# Retiming dynamics of harmonically mode-locked laser solitons in a self-driven optomechanical lattice

Xiaocong Wang,[1,2,3,†] Benhai Wang,[2,†] Wenbin He,[2,*] Xintong Zhang,[1,2,3] Qi Huang,[2] Zhiyuan Huang,[3] Xin Jiang,[2] Philip St.J. Russell[2] and Meng Pang,[1,2,3,4,**]

[1] Department of Optics and Optical Engineering, University of Science and Technology of China, Hefei 230026, China
[2] Russell Centre for Advanced Lightwave Science, Shanghai Institute of Optics and Fine Mechanics and Hangzhou Institute of Optics and Fine Mechanics, Chinese Academy of Sciences, Shanghai 201800, China
[3] State Key Laboratory of High Field Laser Physics and CAS Center for Excellence in Ultra-intense Laser Science, Shanghai Institute of Optics and Fine Mechanics CAS, Shanghai 201800
[4] Hangzhou Institute for Advanced Study, University of Chinese Academy of Sciences, Hangzhou 310024, China
[†] These authors contributed equally to this work.
[*] wenbin.he@r-cals.com
[**] pangmeng@siom.ac.cn

**Abstract:** Harmonic mode-locking, realized actively or passively, is an effective technique for increasing the repetition rate of lasers, with important applications in optical sampling, laser micro-machining and frequency metrology. It is critically important to understand how a harmonically mode-locked pulse train responds to external perturbations and noise, so as to make sure that it is stable and resistant to noise. Here, in a series of carefully designed experiments, we elucidate the retiming dynamics of laser pulses generated in a soliton fiber laser harmonically mode-locked at ~2 GHz to the acoustic resonance in a photonic crystal fiber (PCF) core. We characterize the self-driven optomechanical lattice along the PCF using a homodyne set-up, and reveal that each soliton undergoes damped oscillatory retiming within its trapping potential after an abrupt perturbation. In addition we show, through statistical analysis of the intra-cavity pulse spacing, how the trapping potentials are effective for suppressing timing jitter. The experimental results are well described using a dynamic model including dissipation, which provides valuable insight into the stability and noise performance of optomechanically mode-locked laser systems, and may also be useful for studying complex inter-soliton interactions.

## 1. Introduction

Mode-locked lasers [1] are of great importance in frequency metrology [2, 3], micro-machining [4], optical sampling [5], biomedical imaging [6], and optical information technology [7-10]. As nonlinear dissipative systems, they exhibit a wide range of complex multi-pulse phenomena, for example harmonic mode-locking [11-13], which produces a train of regularly spaced pulses at a repetition rate that is a high multiple of the cavity round-trip frequency. Such high-repetition-rate pulse trains are key in many applications [14-16]. Although external modulation has been used to control harmonic mode-locking [17, 18], and intra-cavity interpulse interactions have been studied [19-22], no comprehensive experimental investigation of the retiming dynamics of multiple pulses in a fiber laser—critical to the stability and noise performance—has been reported to date.

In recent years few-GHz acoustic resonances in the µm-sized core of a photonic crystal fiber (PCF) have been used to passively and stably modelock fiber lasers at a repetition rate that is a high multiple of the round-trip frequency [23-25]. A length of PCF is spliced into a soliton laser cavity [26], resulting in hundreds of equally-spaced solitons in the laser cavity. In the steady state, the GHz-rate pulse train coherently drives the PCF core resonance, which in turn acts back on the soliton pulses, regulating their inter-pulse spacing and creating a stable optomechanical lattice [9]. The result is a simple and robust platform within which complex soliton dynamics can be studied and their statistical features explored [27, 28].

In this paper we present the results of a comprehensive experimental and theoretical investigation into the retiming dynamics of solitons in an optomechanically mode-locked soliton fiber laser. Homodyne measurements are used to measure the profile of the optomechanical lattice in the PCF under conditions of stable high harmonic mode-locking. We then introduce abrupt perturbations to selected solitons riding on the lattice using externally-launched pulses, and investigate the retiming behavior using a phase-space dynamic model. Additionally, we use dispersive time-delay interferometry to measure the inter-pulse timing jitter and explore how the optomechanical lattice suppresses noise.

## 2. Formation of optomechanical pulse trapping potential

A sketch of the high-harmonically mode-locked fiber laser system is shown in Fig. 1(a), together with the gain and loss profiles of individual pulses and the optomechanical lattice that regulates the interpulse spacing. An erbium-doped fiber amplifier (EDFA) provides gain and nonlinear polarization rotation (NPR) produces intensity-dependent loss [29] (see details in Supplement 1). The correct balance between gain and loss is critical for initiating mode-locking [30] and stabilizing the pulse energy, while fiber nonlinearity and anomalous average cavity dispersion results in soliton formation [31]. The soliton sequence coherently drives a PCF core resonance, leading to formation of a sinusoidal acoustic strain wave (the acoustic lattice) travelling at the group velocity of the light [9] (Fig. 1(b)). A single soliton rides within each acoustic cycle, trapped by an effective potential that prevents erratic spacing between adjacent pulses (see Fig. 1(c)). If a soliton is pushed from its stable trapping position, it returns to its original position after oscillating to and fro, damped by gain filtering in the EDFA [32]. This retiming oscillation is inevitably coupled to dissipative effects related to the laser gain and NPR-induced loss. The trapping potential also stabilizes the soliton spacing against noise, which would otherwise cause pulse-timing jitter [33].

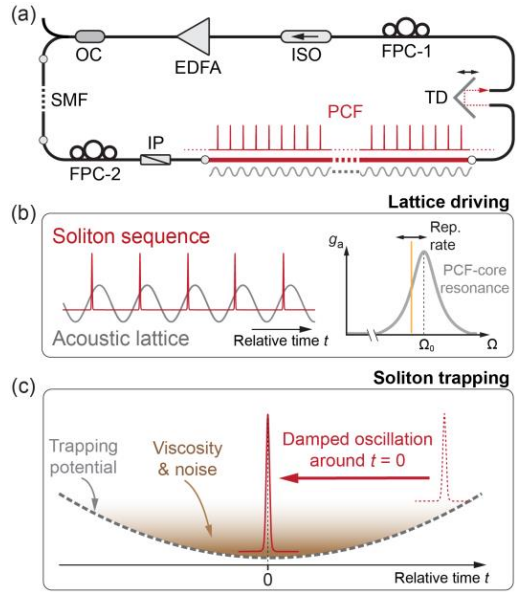

Fig. 1. (a) Sketch of harmonically mode-locked laser cavity based on micro-core PCF and NPR-action. EDFA: erbium-doped fiber amplifier; OC: optical coupler; SMF: single-mode fiber; FPC:

fiber polarization controller; IP, inline polarizer; TD: tunable delay-line; ISO: isolator. (b) The soliton sequence rides on the acoustic field which is coherently driven by the solitons themselves (left) when their repetition rate falls into the acoustic resonance of the PCF-core, forming a self-driven optomechanical lattice. The acoustic gain spectrum (right) has a maximal gain ($g_a$) at the resonance frequency $\Omega_0$ [26]. (c) The soliton in each acoustic cycle is trapped in an effective potential with viscosity and noise, and will take damped oscillatory retiming if deviated from the balanced position.

## 3. Measurement of the autonomous optomechanical lattice

### 3.1 Homodyne system

In the system investigated (Fig. 2(a)) the cavity length was 25 m, and the PCF was 0.7 m long and had a core resonance at 1.876 GHz with a FWHM bandwidth of 15 MHz (see Fig. 2(b)). In the steady state, the laser operated at the 230[th] harmonic, corresponding to a pulse repetition rate of 1.870 GHz and an acoustic wavelength of ~11 cm along the PCF. To probe the acoustic structure along the PCF (i.e. the autonomous optomechanical lattice driven by the soliton sequence), we incorporated a homodyne system into the laser cavity. The output from a single-frequency 1550 nm laser was split into a reference and a probe signal. The probe was launched into the PCF, and the reference and transmitted probe signals were recombined at a 50/50 coupler working as an interferometer that translates the acoustic phase modulation in the PCF into power modulation that can be recorded by a photodetector and an oscilloscope. To linearize the response, the interferometer was operated at quadrature, and both paths were amplified and adjusted using FPCs to maximize the interference contrast. The probe path in the cavity was terminated by an in-line polarizer to avoid perturbing the EDFA.

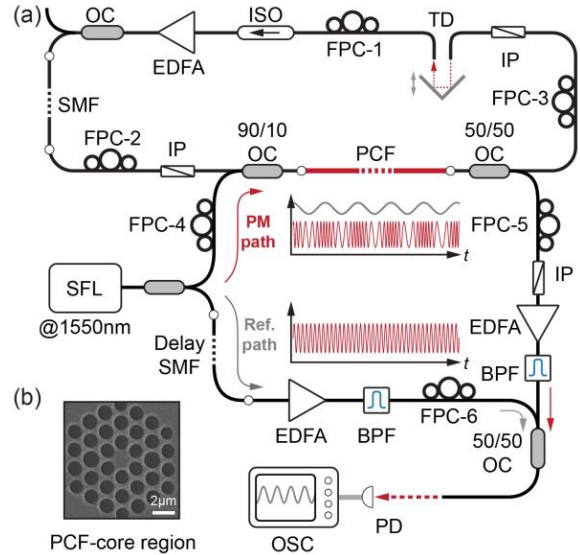

Fig. 2. (a) Sketch of the mode-locked laser, incorporating a Mach-Zehnder interferometer for homodyne measurements. SFL: single-frequency laser at 1550 nm; BPF: band-pass filter, PD: photodetector; OSC: oscilloscope. (b) Scanning electron micrograph (SEM) of the PCF core structure.

### 3.2 Results

The high-harmonic mode-locked laser produced a stable train of pulses at 1562 nm with a fractional bandwidth of 0.0013 and duration 1.3 ps (Fig. 3(a)). The repetition rate was locked to the core resonant frequency, and the optoacoustic gain had a bandwidth of 15 MHz (Fig. 3(b)). The repetition rate could be tuned over ~8 MHz using a variable delay-line (TD), as explained previously [9]. A typical sequence recorded by the oscilloscope, with pulse spacing 535 ps, is shown in Fig. 3(c).

The measured homodyne signal at a repetition rate detuning of 4.9 MHz is sinusoidal in shape, with a single narrow feature (corresponding to a pulse) riding on it in each cycle (Fig. 3(d)). The signal is the result of elasto-optic and Kerr-related refractive index changes, and clearly confirms that the acoustic wave serves as an optomechanical lattice regulating the inter-pulse spacing. When the repetition rate was detuned further to 10.9 MHz, the measured amplitude of the acoustic modulation (lower part of Fig. 3(d)) decreased by ~25%, in good agreement with the predictions of a previous simplified model of the system [9].

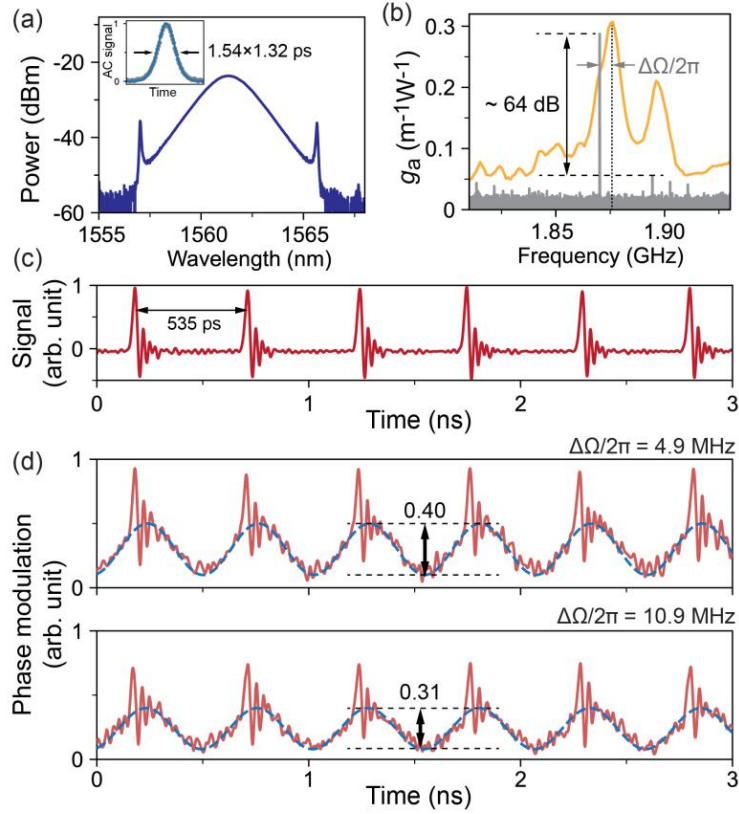

Fig. 3. (a) The optical spectrum and autocorrelation trace (inset) of the output laser solitons. (b) The measured optoacoustic gain in the PCF (orange) [26] and the power spectrum of the output sequence (gray) with a strong spike at the repetition rate. (c) The output pulse sequence recorded by the oscilloscope. (d) The homodyne power signals normalized to a reference power measured at repetition rate detunings ($\Delta\Omega/2\pi$) of 4.9 MHz and 10.9 MHz.

## 4. Oscillatory retiming of laser solitons in the trapping potential

### 4.1 Abrupt perturbations using externally-launched control pulses

To explore the retiming dynamics, we abruptly perturbed a single pulse in the mode-locked sequence and observed its response. Since all the other pulses are undisturbed, the overall structure of the optomechanical lattice is preserved. The perturbation was introduced by externally-launched control pulses generated using an electro-optical modulator and spaced by the round-trip time of the laser cavity (Fig. 4(a)). To avoid disturbing the EDFA, the control pulses were eliminated from the cavity after the SMF using an intra-cavity polarizer (IP).

As the control pulses travel along the 12-m-long SMF, cross-phase modulation causes the selected soliton to move away from its equilibrium position in the trapping potential (see details in Supplement 1). This shift in temporal position is accompanied by changes in group velocity and pulse energy (Fig. 4(b)). The perturbed pulse train is extracted from the cavity at output port OC-2 (Fig. 4(a)) and viewed using a fast oscilloscope. When the control pulses are turned off, the perturbed soliton undergoes damped oscillations in relative time, ultimately returning to its equilibrium position (Fig. 4(c)).

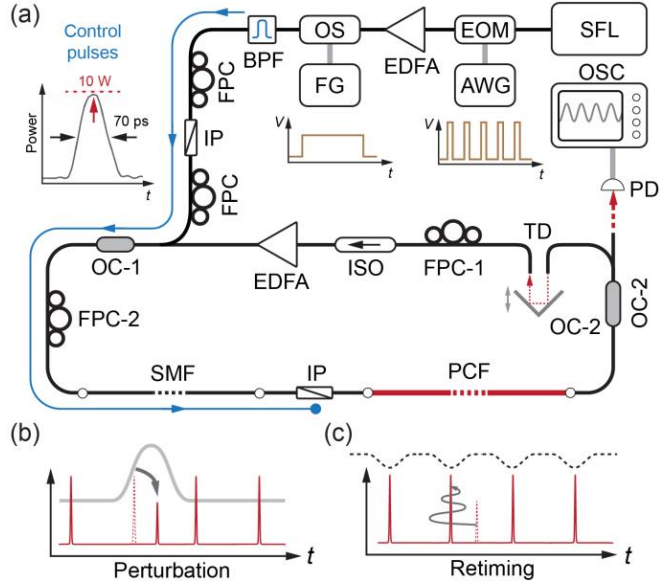

Fig. 4. (a) Sketch of the mode-locked cavity including an external system for launching control pulses to perturb individual laser pulses. The blue line marks the path of the control pulses. (Top-left inset: temporal profile of the external pulses.) EOM: electro-optical modulator; AWG: arbitrary waveform generator; OS: optical switch; FG: function generator. (b) Temporal shift of a selected soliton in response to the control pulse. (c) Damped retiming oscillation of the soliton in its trap after the perturbation is switched off.

*4.2 Observations of damped retiming oscillations*

Three measurements of the retiming dynamics are plotted in Fig. 5(a) in a time-frame moving at the average cavity group velocity. A key result is that the soliton energy falls and the lifetime of the damped retiming oscillations increases as the perturbation strength rises (Fig. 5(b)). In Case #1, a ~25% decrease in pulse energy is observed, and the 1/e damping time is ~1 ms. In Case #2, the perturbation is stronger, the pulse energy decreases by ~50%, and the damping time is ~6.5 ms, corresponding to ~4 cycles. In both these cases the pulse energy recovers to its initial value. In Case #3 the perturbation strength is further increased, causing an ~80% decrease in soliton energy. In this case the soliton is no longer able to recover to its initial energy, and intriguingly the retiming oscillations have very low damping, persisting even after the soliton has become very weak, ultimately vanishing into the noise background. In all these cases, the signals from the two unperturbed next-neighbor solitons are included for comparison, confirming that the optomechanical lattice remains stable during the measurements. In addition, the period of the retiming oscillations is determined by the trapping potential in the acoustic lattice and could be varied by adjusting the cavity parameters, in agreement with previous work [9]. (See results in Supplement 1.)

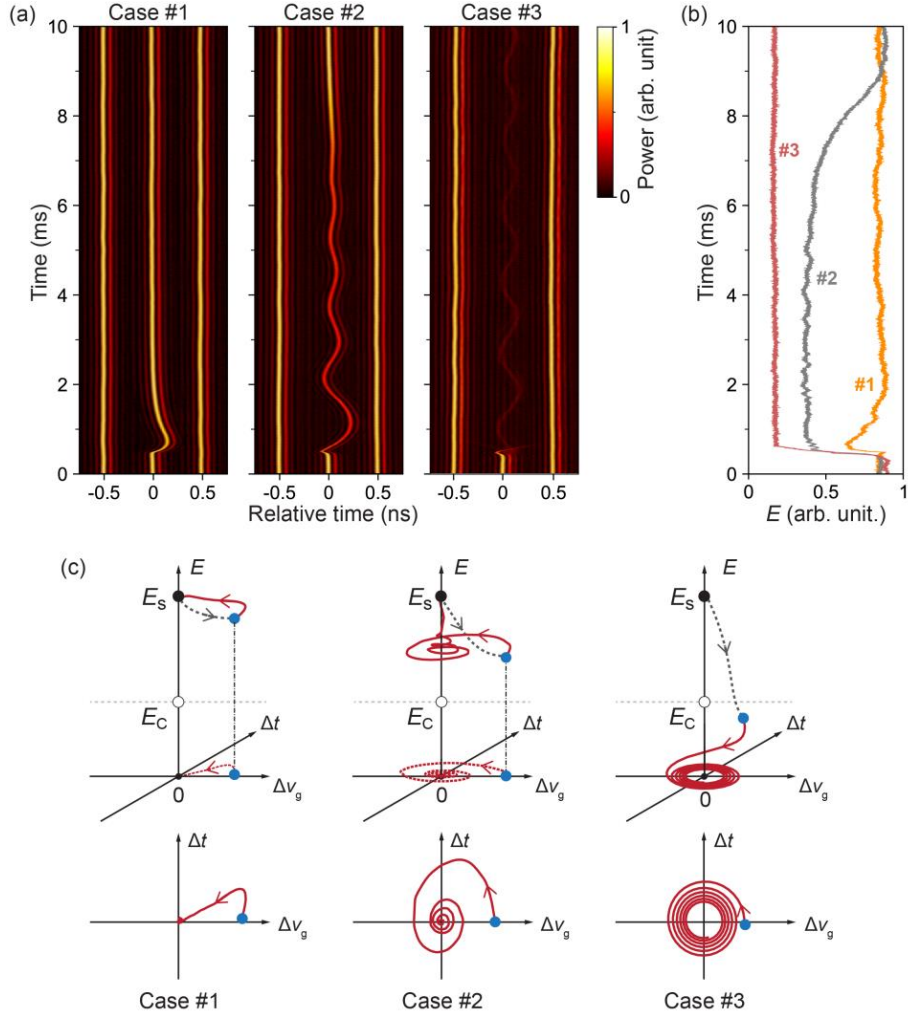

Fig. 5. (a) Observed retiming dynamics of individual solitons after perturbation by control pulses (the perturbation strength increases from Case #1 to Case #3). Note that the two next-neighbor solitons remain unperturbed. (b) Evolution of the soliton energy and (c) phase-space trajectories (red-curves) for the three cases in (a). The dashed gray curves, starting at the black dot and ending at the red dot, mark the abrupt change in soliton state induced by the perturbation.

## 4.3 Phase-space description and dissipative model

The mechanism underlying this damped oscillatory behavior can be best understood using a phase-space description involving the pulse energy ($E$), and the deviations $\Delta\tau$ in time and $\Delta v_g$ in group velocity from the equilibrium values. The measured dynamics (Fig. 5(a)) can then be mapped on to $E$-$\Delta\tau$-$\Delta v_g$ space (Fig. 5(c)). The unperturbed soliton exists at a fixed-point attractor on the $E$-axis, with $\Delta\tau = 0$ and $\Delta v_g = 0$ (black dots in Fig. 5(c)). For a small decrease in soliton energy (Case #1), the trajectory is heavily damped and the system moves nearly monotonically towards the attractor from the perturbed point (blue dot). For a stronger perturbation, the soliton energy decreases significantly, and the system follows a spiral trajectory towards the attractor. For a sufficiently strong perturbation, the soliton energy falls below a critical value $E_c$ (Case #3), and the system is no longer captured by the attractor, moving instead along a spiral trajectory towards the zero-energy plane, where the pulse disappears.

Following the well-known approach by Haus [1], we have developed a dynamic analytical model incorporating both retiming [9] (Fig. 1(c)) and changes in dissipation caused by laser gain and nonlinear polarization rotation (see Fig. 6(a)). The pulse energy $E$ is governed by:

$$\frac{dE}{d\tau} = g(B_s)E - \alpha(P_s)E \approx \big(g(k_s E) - \alpha_0 \sin(k_\alpha E^2)\big)E \tag{1}$$

where $\tau = T/T_R$ is the normalized time, $T_R$ the cavity round-trip time, $g(B_s) \approx g(k_s E)$ the laser gain (which decreases as the soliton bandwidth $B_s$ increases (at larger $E$) due to gain filtering [32]), and $\alpha(P_s)$ is the loss induced by nonlinear polarization rotation, which depends sinusoidally on the peak pulse power $P_s \approx E/\tau_p \approx k_\alpha E^2$, where $\tau_p$ is the pulse duration and $k_\alpha$ a heuristic constant [29]. As shown in the sketch in Fig. 6(a), the combined gain term in Eq. (1) oscillates around zero with increasing $E$. The first zero at $E = E_C$ has positive slope and is unstable; for $E < E_C$ the pulse energy decays to zero, and for $E > E_C$ it increases until it reaches the second zero at $E = E_S$, which has a negative slope and is therefore stable; the soliton energy is then clamped at this value. At larger energies the laser gain falls off, so that there are no more zeroes in the combined gain, and no further stable points. (See details in Supplement 1.)

The temporal shift in pulse position $\Delta\tau$ is directly related to the change in group velocity $\Delta v_g$:

$$\frac{d\Delta\tau}{d\tau} = -\frac{\Delta v_g}{v_{g0}} \quad , \tag{2}$$

where $v_{g0}$ is the group velocity at the equilibrium point. Defining the normalized group-velocity change as $\Delta\hat{v}_g = \Delta v_g/v_{g0}$, the time dependence of $\Delta\hat{v}_g$ is governed by:

$$\frac{d(\Delta\hat{v}_g)}{d\tau} = -\Gamma(B_s)\Delta\hat{v}_g + F(\Delta\tau) \quad , \tag{3}$$

where $\Gamma(B_s)$ is the damping rate, and $F(\Delta\tau)$ is the "restoring force", which is linearly proportional to $\Delta\tau$, assuming a quadratic trapping potential.

Equations (2)&(3) can be combined to yield a 2$^{nd}$-order differential equation:

$$\frac{d^2\Delta\tau}{d\tau^2} + \Gamma(B_s)\frac{d\Delta\tau}{d\tau} - F(\Delta\tau) = 0 \quad , \tag{4}$$

which describes the behavior of the system when perturbed from its equilibrium position. A key point here is that $\Gamma(B_s)$ increases with soliton bandwidth due to gain filtering [32], so that lower soliton energies lead to narrower pulse bandwidths and slower damping rates. (See details in Supplement 1.)

The retiming dynamics described by this analytical model are summarized in Fig. 6(b) with three typical cases. In Case #1 the soliton energy after perturbation is only slightly lower than $E_s$, and the system response is overdamped, quickly returning to equilibrium. In Case #2 the soliton energy after perturbation is significantly lower than $E_s$ but still above $E_c$, and the system oscillates with a longer damping time, returning to equilibrium more slowly. In Case #3 post-perturbation soliton energy is lower than $E_c$, and thereafter slowly falls, exhibiting long-lived oscillations that continue until the soliton has almost vanished. The numerical results based on this dynamic model are given in Fig. 6(c), showing damped oscillations around $E_s$ for $E > E_C$ and long-lived oscillations for $E < E_C$ that decay slowly towards zero energy, which is in good agreement with the experimental results in Fig. 5.

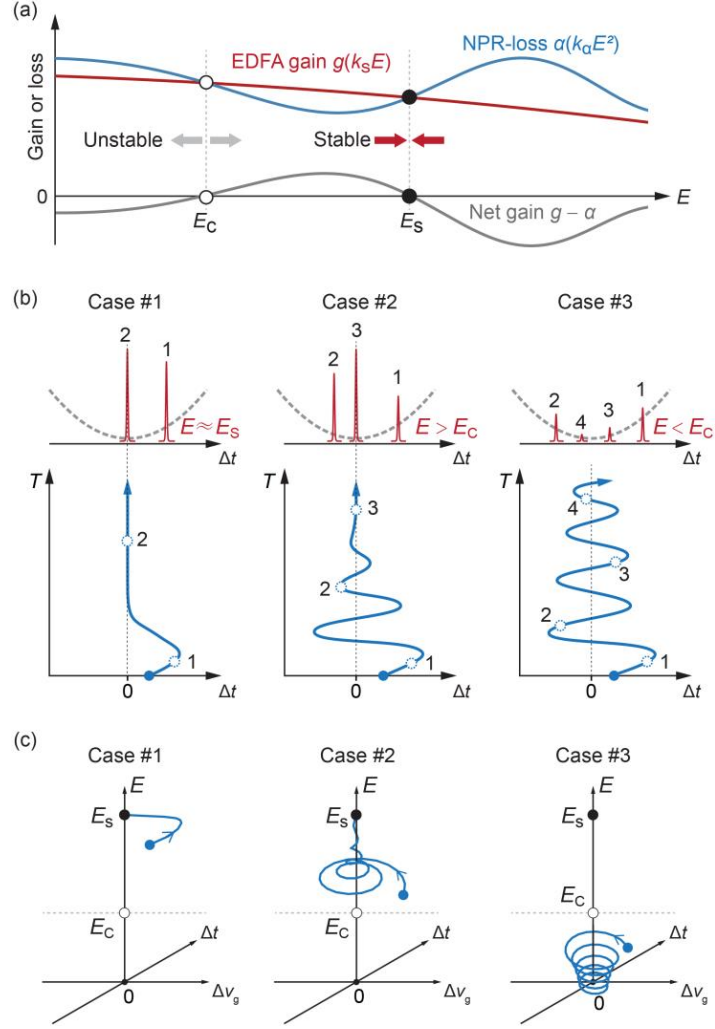

Fig. 6. (a) Intensity-dependent NPR-induced loss and energy-saturable EDFA gain leads to a stable pulse energy $E_s$, while a critical pulse energy $E_c$ exists below which the gradually disappears. (b) Illustration of the key features of the retiming dynamics, for three different initial conditions. (c). The simulated phase-space trajectories based on the dynamic model described by Eqs. (1) – (3) for the retiming processes. Three different initial conditions are chosen corresponding to (b).

## 5. Suppression of inter-pulse jitter in the trapping potential

### 5.1 Dispersive time-delayed interferometry

In the absence of external perturbations, the pulse timing will be influenced by noise sources such as amplified spontaneous emission (ASE) [33], resulting in unpredictable inter-pulse jitter. At the same time, the acoustic wave in the PCF, driven by the intra-cavity pulse train, can in principle suppress the timing jitter by continuously enforcing a regular pulse spacing. To probe the inter-pulse jitter, we employed dispersive time-delayed interferometry (DTDI) [34] to follow real-time fluctuations in inter-pulse spacing, which are too fast to be resolved by direct detection using fast electronics. In the DTDI set-up, the output pulse sequence from the mode-locked laser is divided into two paths, one being temporally delayed before being recombined with the other path (Fig. 7(a)). The result is a train of closely spaced pulse pairs (Fig. 7(b)), which can be resolved after time-stretched dispersive Fourier transformation (TS-DFT) over a 5 km length of SMF [35]. The residual pulse spacing is retrieved from the fringe period of the resulting spectral interferogram, with a temporal resolution less than 1 ps [36] ( Fig. 7(c)). Jitter-induced changes in inter-pulse spacing can then be monitored by comparing the fringe periods for each pulse pair.

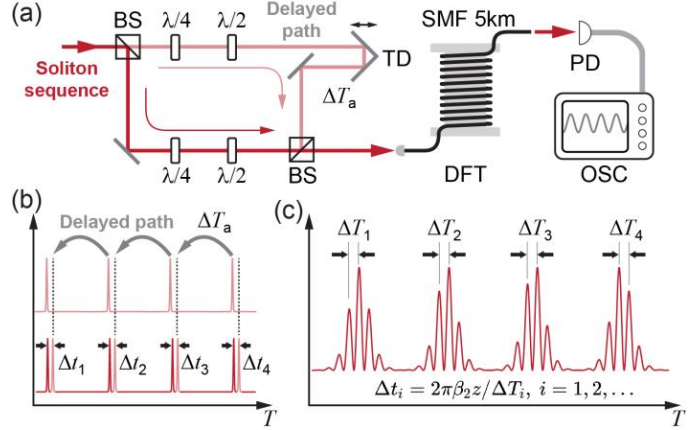

Fig. 7. (a) The DTDI set-up for resolving fluctuations in inter-pulse spacing. BS: beam splitter. (b) The pulse train is split into two paths, one of which has a variable delay line. The two paths are then recombined at a beam-splitter, and the delay adjusted until there is a train of closely-spaced pulse pairs. (c) TS-DFT signal of the pulse-pair sequence, in which the temporal spacing between the fringes is inversely proportional to the temporal spacing of each pulse pair.

*5.2 Statistical analysis of inter-pulse jitter*

The DTDI signal was recorded using a fast oscilloscope, resulting in 230 real-time sequences of temporal interferograms formed by neighboring solitons. The temporal interferogram of a randomly selected pulse-pair is plotted over consecutive round-trips in Fig. 8(a). It is clear that the resulting interferograms generally remain stable in the short term (~50 μs), while exhibiting prominent random drifts over longer times (~15 ms). The long-term pulse spacing is plotted in the lower part of Fig. 8(b), showing a bounded stochastic oscillation around an average position. The retrieved relative phase between consecutive pulses, in contrast, exhibits an unbounded random-walk (Fig. 8(b)), indicating that the carrier phase of the solitons are uncorrelated [37]. Analyzing the distribution probability of the retrieved spacings for all 230 pulse pairs over a time-scale of 5 ms resulted in a narrowly-distributed Gaussian profile centered at the average position, as shown in Fig. 8(c). Most importantly, this distribution widens when the repetition rate is detuned from the acoustic resonant frequency, as shown in Fig. 8(c), where the full-width-half-maximum jitter distribution broadened from 4.8 ps to 9.2 ps as the frequency detuning was changed from 5 MHz to 12.3 MHz. Note that the center of the distribution shifts slightly due to the change of repetition rate. These results confirm that the optomechanical lattice suppresses timing jitter, with an effectiveness that falls off as the repetition rate is detuned from the acoustic resonance. This is because weaker driving of the acoustic wave (see upper part of Fig. 3(d)) weakens the trapping potential, reducing suppression of inter-pulse jitter.

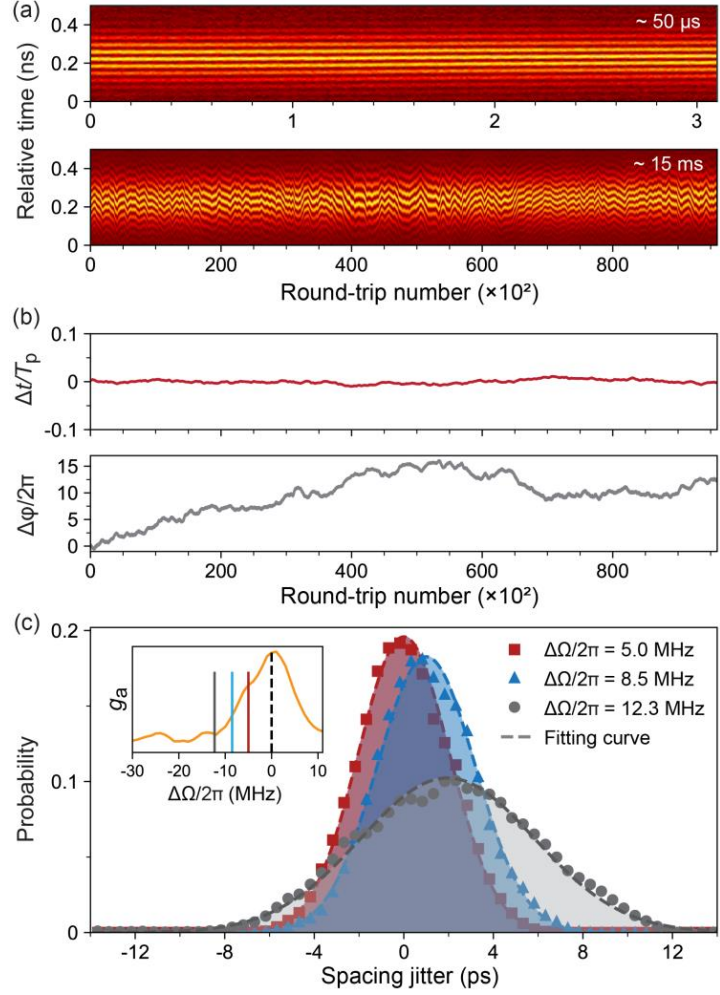

Fig. 8. (a) The evolution over 15.5 ms of the TS-DFT signal for an arbitrarily selected pulse-pair over consecutive round-trips (zoom-in over 50 µs is also shown). (b) Upper: the retrieved spacing jitter ($\Delta\tau$) from (a), relative to the average inter-pulse spacing $T_\text{p}$ in the pulse train as it leaves the laser. Lower: the relative phase ($\Delta\varphi$) between the pulse pair. (c) The probability distribution of the inter-pulse jitter for three different values of repetition rate detuning. Inset: the optoacoustic gain profile, with the three values of detuning marked.

## 6. Discussion and conclusions

Self-stabilization of a high harmonic mode-locked pulse train can be achieved by strong optoacoustic interactions in a small-core PCF, resulting in the generation of an acoustic wave that, close to cut-off, travels at the group velocity of the light and forms an optomechanical lattice that clamps each laser soliton into a trapping potential. In the presence of external perturbations and intrinsic noise, the optomechanical lattice robustly enforces the pulse repetition rate, stabilizing the inter-pulse spacing.

An open question is how the harmonic mode-locking gets started. In the experiments, mode-locking self-initiates from noise without any ordering, whereas the steady-state is a highly-ordered structure with stable interactions between a regular sequence of solitons and a coherently-driven autonomous optomechanical lattice. The retiming dynamics of an individual soliton in the steady state cannot be used to explain how a highly ordered sequence of hundreds of solitons is established [38]. A further interesting question concerns the carrier-wave phase relationship between pulses when the laser is stably mode-locked [37] (lower part of Fig. 8(b)). It remains unclear how this is affected by inter-soliton interactions, or whether it can be manipulated through cavity control. The frequency-domain (comb) structure of the laser is also unclear due to the massive number of solitons and their complicated collective behavior. From another perspective, the complexity of the optoacoustically modelocked laser system may provide rich possibilities for manipulating the time- and frequency-domain structures of high-repetition-rate lasers.[39-41]

The reported laser system provides an elegant platform for studying multi-pulse interactions and their dynamics, enriching our understanding of the retiming dynamics of optoacoustically mode-locked fiber lasers, and laying the ground-work for future experimental and theoretical studies of self-stability and noise-suppression in such lasers.

**Funding.** National Natural Science Foundation of China (Grant No. 62375275 and 62275254); Strategic Priority Research Program of the Chinese Academy of Science (XDB0650000); Shanghai Science and Technology Plan Project Funding (Grant No.23JC1410100); Fuyang High-level Talent Group Project.
**Disclosures.** The authors declare that there are no conflicts of interest related to this article.
**Data availability.** The data underlying the results presented in this paper is not publicly available at this time but may be obtained from the authors upon reasonable request.
**Supplemental document.** See Supplement 1 for supporting content.

# Supplement for "Retiming dynamics of harmonically mode-locked laser solitons in a self-driven optomechanical lattice"

XIAOCONG WANG,[1,2,3][†] BENHAI WANG,[2,†] WENBIN HE,[2,*] XINTONG ZHANG,[1,2,3] QI HUANG,[2] ZHIYUAN HUANG,[3] XIN JIANG,[2] PHILIP ST.J. RUSSELL[2] AND MENG PANG[1,2,3,4,**]

[1]Department of Optics and Optical Engineering, University of Science and Technology of China, Hefei 230026, China
[2]Russell Centre for Advanced Lightwave Science, Shanghai Institute of Optics and Fine Mechanics and Hangzhou Institute of Optics and Fine Mechanics, Chinese Academy of Sciences, Shanghai 201800, China
[3]State Key Laboratory of High Field Laser Physics and CAS Center for Excellence in Ultra-intense Laser Science, Shanghai Institute of Optics and Fine Mechanics CAS, Shanghai 201800
[4]Hangzhou Institute for Advanced Study, University of Chinese Academy of Sciences, Hangzhou 310024, China
[†]These authors contributed equally to this work.
[*]wenbin.he@r-cals.com
[**]pangmeng@siom.ac.cn

This document provides supplementary materials to "Retiming dynamics of harmonically mode-locked laser solitons in a self-driven optomechanical lattice", with more details on the experimental setups, the observations and the theoretical model.

## 1. Optoacoustically mode-locked laser: setup details.

The optomechanically mode-locked soliton fiber laser shown in Fig. 1(a) of the main text employed a 1-m-long erbium-doped fiber with a peak absorption of 30 dB/m at 1530 nm as the gain fiber. The gain fiber is pumped by two laser diodes at 980 nm with a combined pump power of 1 W. A tunable delay-line (TD) is inserted to ensure that one harmonic of the cavity round-trip frequency would fall into the acoustic resonance of the solid-core PCF[1]. Two fiber polarization controllers (FPCs) are used inside the laser cavity for adjusting the polarization bias to enable the self-starting of the laser mode locking through nonlinear polarization rotation (NPR) [2]. A third FPC is inserted between polarizer and the PCF in order to launch the linearly polarized laser light into one principal axis of the PCF[1]. This particular FPC was fixed in all the subsequent experiments and therefore we have omitted it in all the setup sketches in the main text (Figs. 1, 2, and 4). By carefully adjusting the FPCs, the laser could be arranged to mode-lock at the 230[th] harmonic of the cavity round-trip frequency. The output soliton sequence is sent to a diagnostic set-up which consists of two ports. At one port the optical spectrum of the soliton pulses (see Fig. 3(a)) was recorded directly using an optical spectrum analyzer (OSA, Yokogawa, AQ6374), with a spectral resolution of 0.02 nm. At the other port, the time-domain sequence of the solitons (as shown in Fig. 3(c)) was recorded using a fast photo-detector (PD) and an oscilloscope (OSC, Tektronix, DPO75902SX), with a sampling rate of 50 GSa/s and a detection bandwidth of 15 GHz.

## 2. Perturbation dynamics of control pulses

The control pulses we used for perturbing the intra-cavity solitons are generated by modulating a 1550-nm single frequency CW laser (NKT Koheras, linewidth <100 Hz) using an electro-optical modulator, as shown in Fig. 4 in the main text. The duration of externally launched pulses is 70 ps, which is ~50 times wider than soliton, ensuring sufficient overlapping between the control pulses and the perturbed solitons when they co-propagate in the SMF. The control pulses were generated at intervals exactly corresponding to the cavity round-trip time of the mode-locked lasers, in order to perturb the selected soliton repeatedly. The control pulses were amplified to gain a peak power of ~10 W so as to ensure the strength of XPM effect with the solitons. An optical switch was used to set a 200-μs time window for the perturbation (with a rise/fall time of 300 ns). Two FPCs and an inline polarizer was used to adjust the polarization states of the control pulses in order to eliminate the control pulses at the inline polarizer in the mode-locked cavity and thus avoiding further perturbation in the EDFA section.

The perturbation exerted by the control pulses not only induced deviations in the position and group velocity of the perturbed soliton, but also leads to temporary attenuation of the soliton energy, as already demonstrated in our previous work [3, 4]. Due to the properties of laser solitons[1], the decrease in its energy would cause spectral bandwidth narrowing simultaneously, which would then affect the damping strength of their retiming oscillation in the optomechanical lattice. The spectral bandwidth narrowing of the perturbed solitons can be revealed using the DFT method, which linearly maps the profile of their optical spectrum into the time domain [5]. We used a 5-km-long SMF for dispersive stretching of the output soliton sequence at OC-2. Two examples are provided below, as shown in Fig. S1. In the first example (Fig. S1(a)), the control pulses overlapped with the solitons with a slight discrepancy in the repetition rate, dragging the soliton away from the balanced position [3], meanwhile causing some decrease in the energy (above the critical energy $E_c$) [4]. The DFT signal (shown in Fig. S1(b)) clearly shows the narrowing of the soliton spectrum after the perturbation, although both the energy and bandwidth recovered later during the retiming of the solitons. In the second example (Fig. S1(c)), the control pulses exerted stronger perturbation upon the selected soliton, causing significant reduction in the soliton energy (below $E_c$). In this case the soliton can no longer recover, while we can see from the DFT signal that the soliton bandwidth kept narrowing as the soliton gradually vanished. Note that Fig. S1 (a) and (c) are recorded at OC-1 in order to show the relative position of control pulses and the perturbed solitons. Meanwhile the Fig. S1(b) and (d) are recorded at OC-2 before which the control pulses have already been eliminated in order to clearly show the DFT signal of the perturbed soliton.

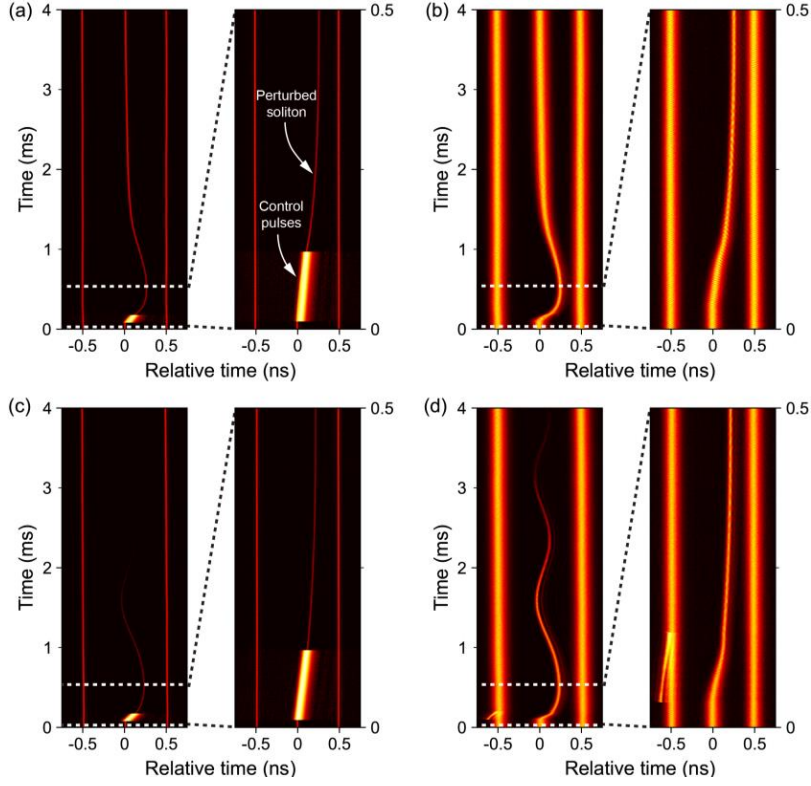

Fig. S1. (a) Perturbation and subsequent over-damped oscillatory retiming of the selected soliton recorded at OC-1. The perturbation region is zoomed-in in the right panel for clear illustration of the relative position between the control pulses and the perturbed soliton. Note that the two next-neighbor solitons remain unperturbed. (b) The DFT signal corresponding to (a) recorded at OC-2. (c) Perturbation by control pulses was increased, which caused vanishing of the perturbed soliton during the oscillatory retiming. (d) The DFT signal corresponding to (c) recorded at OC-2. Note that slight leakage of the control pulses was observed in the DFT signal, while the leakage did not cause obvious perturbations upon the other solitons in the cavity.

## 3. The period of retiming oscillations

The period of the retiming oscillations of the perturbed soliton is determined by the restoring "force" induced by the trapping potential, which can be varied under different cavity parameters. According to Eq. S12 in Ref.[4], under linear approximation, the restoring force $F(\Delta\tau)$ given in Eq.(4) in the main text is related to the temporal deviation $\Delta\tau$ according to:

$$F(\Delta\tau) = -\frac{\omega_0 \Delta n'' L_{\text{PCF}} \beta_2^{\text{ave}} L_R}{c} \Delta\tau \quad , \tag{S1}$$

in which $\Delta\tau$ is the temporal deviation of the soliton from the balanced position normalized to the cavity round-trip time $T_R$ (i.e. $\Delta\tau = (t-t_0)/T_R$ where $t_0$ is the balanced position), $\omega_0$ the carrier frequency, $L_{\text{PCF}}$ the PCF length, $\beta_2^{\text{ave}}$ the average cavity GVD (which is a negative value in unit of ps$^2$/km), $L_R$ the cavity length, and $c$ the vacuum light speed, $\Delta n''$ is the second-order Taylor coefficient of the index modulation induced by the acoustic vibration in the PCF at the balanced position of the laser soliton. The Taylor expansion of the index modulation in real time scale about the balanced position up to the second order term can be written as below:

$$\Delta n(t-t_0) = \Delta n_0 + \Delta n'(t-t_0) + \frac{1}{2}\Delta n''(t-t_0)^2 \quad , \tag{S2}$$

in which $\Delta n_0$, $\Delta n'$, and $\Delta n''$ are the Taylor coefficients for the expansion terms from the zeroth to the second order ($\Delta n''$ has a unit of $\text{sec}^{-2}$). Stable trapping of the soliton in the acoustic lattice requires that $\Delta n'' < 0$. As a result, the repetition rate is locked at a frequency lower than the resonance frequency, typically at the FWHM position of the acoustic resonance [4].

The dimensionless coefficient of $\Delta\tau$ in the right-hand side of Eq. (S1) is simply the squared angular frequency of the retiming oscillation enforced by the trapping potential. We can then roughly estimate the corresponding oscillation period using the following practical numbers. We can calculate that $\Delta n''$ is in the order of $-10^{-13}$ $\text{ps}^{-2}$ given that a 2-GHz vibration in the PCF-core leads to a $\Delta n_0$ in the order of $10^{-8}$ along the PCF [6]; the total group delay $\beta_2^{\text{ave}} L_R \approx -0.5$ $\text{ps}^2$ ($\beta_2^{\text{ave}} \approx -20$ $\text{ps}^2/\text{km}$ and $L_R \approx 25$ m), and $\omega_0/c = 2\pi/(1.55\mu m) \approx 4$ $\mu m^{-1}$. Using the expression of restoring force given in Eq. (S1), we can estimate that the oscillation period is $\sim 10^4$ in normalized time scale $\tau$ with respect to the round-trip time ($T_R \approx 100$ ns for 25-m-length cavity), which is roughly in the order of 1 ms in the real time scale, agreeing well with our experimental observation.

The oscillation period can be varied under different cavity configurations, as indicated by Eq. (S1), which have also been verified in our experiments. Two examples are provided in Fig. S2 which gives the retiming oscillations under two different cavity configurations. In the first example (Fig. S2(a)), we have $L_{\text{PCF}} = 0.7$ m and $L_R = 24$ m, and we can observe that the retiming oscillation lasted for four cycles over ~6.5 ms, giving an oscillation period of ~1.6 ms. In the second example (Fig. S2(b)) we have increased the fiber length such that $L_{\text{PCF}} = 1.5$ m and $L_R = 35$ m, (thus $\beta_2^{\text{ave}}$ remained almost unvaried). We can notice that due increase of the PCF length and the cavity length, the restoring "force" is enhanced and thus oscillation period has been reduced significantly to ~ 1.2 ms (four cycles of oscillations are observed over a time span of ~ 4.6 ms). According the Eq. (S1), if we assume that $\Delta n''$ is unvaried, the scaling in the oscillation period should be about 1.21 times (1.77 times in normalized time scale, while $T_R$ are different in these two cases by 1.46 times), which agrees quite well with the experimental results that yielded a scaling of 1.33 times (from 1.2 to 1.6 ms). The scaling in oscillation period does not fully match the change in the fiber length probably due to the slight difference of $\Delta n''$ in these two cases, which is sensitive to both the magnitude of the acoustic vibration and the frequency detuning of the repetition rate, and is meanwhile difficult to characterize precisely in experiments.

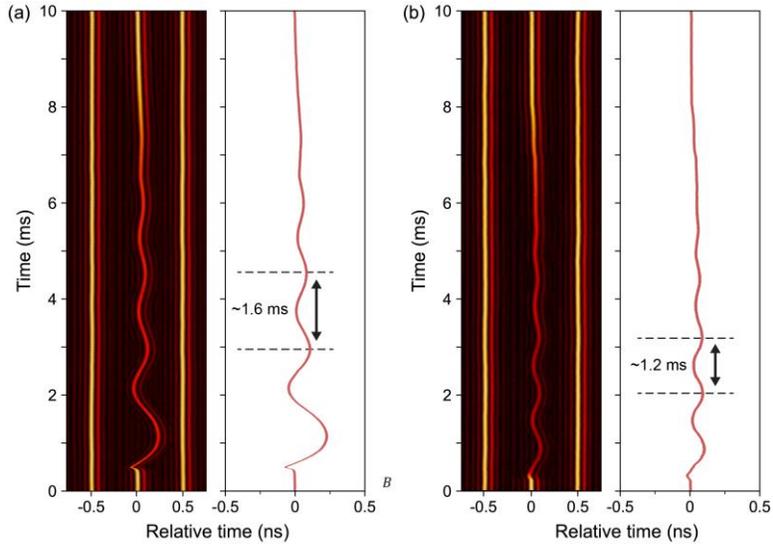

Fig. S2. Retiming oscillations observed in two different cavity configurations. (a) The cavity includes a 0.7-m-long PCF and has a 24-m-long cavity length. The observed retiming oscillation period is ~1.6 ms. The trajectory of the perturbed soliton is extracted and plotted separately on the right panel for clear illustration. (b) The cavity includes a 1.5-m-long PCF and has a 35-m-long cavity length. The observed retiming oscillation period is ~1.2 ms.

In fact, due to the sinusoidal profile of the acoustic lattice, the restoring force induced by the acoustic lattice becomes nonlinearly related to the position deviation $\Delta\tau$ given large initial deviations. The restoring force around the balanced position can then be expressed using Taylor expansion as:

$$F(\Delta\tau) = k_1 \Delta\tau + k_2 \Delta\tau^2 + k_3 \Delta\tau^3 + \cdots \quad , \tag{S3}$$

Due to the nonlinear terms in $F(\Delta\tau)$, the trajectory of retiming oscillation in the time domain observed in our experiments are slightly different from a damped sinusoidal trajectory. We have revealed from numerical calculations that due to the existence of these nonlinear terms, the profile of the first cycle in the retiming oscillation could be sensitive to the initial deviations (position and velocity). As the oscillation quickly damped to the close vicinity of the balanced position, the trajectory can then be well described using a harmonic potential assumption. Further investigations are needed to gain better insights into the nonlinear properties of the restoring forces induced by the acoustic lattice.

## 4. Dissipative model of retiming oscillation

We have developed a phase-space description and a corresponding dissipative model (Eqs. (1) – (4)) that describe the retiming oscillation of the laser soliton. The key mechanism in this model is that the presence of a trapping potential leads retiming oscillation of deviated solitons, while the damping strength of the retiming oscillation is determined by the soliton bandwidth through gain filtering effect. The soliton bandwidth is directly proportional to the soliton energy due to the soliton-area theorem [1] which implies $E \propto 1/\tau \propto B_s$. Meanwhile the dynamics of soliton energy is governed by the balance between the EDFA gain and the NPR-induced loss as described Eq. (1) in the main text. In this way, the retiming dynamics are coupled to the dissipative dynamics of the intra-cavity solitons. We employ the following assumption concerning the dissipative terms in Eq. (1) for the numerical simulations given in Fig. 6(c) in the main text. Firstly, the EDFA gain $g$ is related to the soliton bandwidth $B_s$ (and thus to the soliton energy $E$) through gain filtering effect according to:

$$g(B_s) = g_0\left(1 - \frac{B_s^2}{\Omega_g^2}\right) = g_0\left(1 - \frac{(k_s E)^2}{\Omega_g^2}\right) \tag{S4}$$

where we assumed a quadratic dependence of the gain coefficient upon the soliton bandwidth [7] with $g_0$ being the gain for quasi-CW pulse and $\Omega_g$ is the gain bandwidth, and $k_s$ a heuristic constant. This assumption is valid given that the soliton bandwidth is small compared with the gain filtering bandwidth. Note that the gain saturation effect can also lead to lower gain for higher soliton energy, which can further decrease the gain coefficient when the soliton energy is increased.

Secondly, the NPR-induced loss is related to the peak power of the soliton $P_s$ (and thus the soliton energy $E$ since $P_s \propto E^2$) sinusoidally, and the complete form of $\alpha(P_s)$ used in numerical simulation is given by [2]:

$$\alpha(P_s) = \alpha(k_\alpha E^2) = \alpha_1 + \alpha_0 \sin(k_\alpha E^2 + b_\alpha) \tag{S5}$$

in which $\alpha_1$ is the unsaturable loss, $\alpha_0$ is a constant related to the magnitude of the NPR-induced loss, $k_\alpha$ indicates the sensitivity of the NPR-loss to the soliton energy, $b_\alpha$ indicates the polarization bias of the cavity, which needs to be adjusted to form proper values of $E_s$ and $E_c$ as shown in Fig. 6(a) in the main text.

At last, the damping strength $\Gamma$ is affected by $B_s$ due to the gain filtering effect. This effect can be illustrated using the conceptual sketch in Fig. S3. The initial change of the group velocity $\Delta v_g$ of the perturbed soliton can regarded as the consequence of a change in the carrier frequency $\Delta \omega$ of the soliton following $\Delta v_g = -\beta_2^{\text{ave}} v_{g0}^2 \Delta \omega$ ($v_{g0}$ is the unperturbed soliton group velocity). The gain filtering effect provided by the EDFA tends to eliminate $\Delta \omega$ by pulling the center of the soliton spectrum back to that of the gain spectrum, as conceptually illustrated in Fig. S3(a). During this process, $\Delta v_g$ is also decreased, leading to an effective damping for the oscillation motion of the soliton in the trapping potential (See Fig. S3).

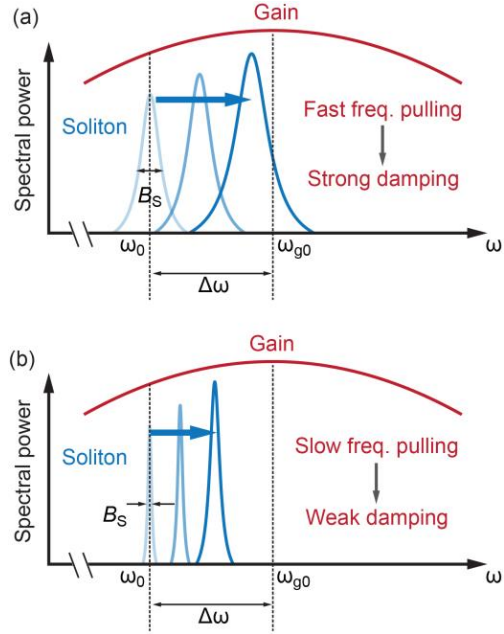

Fig.S3. Conceptual illustration of the bandwidth-dependent damping strength for soliton retiming. (a) Soliton spectrum with a bandwidth of $B_s$ and a shifted carrier frequency $\omega_0$ away from the from the spectral center of the EDFA gain $\omega_{g0}$ by $\Delta \omega$. The gain filtering effect tends to pull the center of the soliton spectrum back to the center of the gain by eliminating $\Delta \omega$, which causes the damping of the soliton motion. (b) Given a narrower soliton bandwidth, the pulling of the soliton spectrum is much slower with each amplification per round-trip, leading to weaker damping compared to (a). Note during the amplification the soliton energy gradually increase, leading to broader bandwidth.

As a consequence, we can expect that solitons with larger bandwidth would be more sensitive to the unbalanced gain profile away from the spectral center of the gain, and thus would be damped more strongly during the retiming oscillation. As a simple approximation, we assume the damping strength depends quadratically upon $B_s$ (and thus $E$) as:

$$\Gamma(B_s) = k_\Gamma E^2 \quad , \tag{S6}$$

As a results, a much lower damping strength would be expected if the soliton energy is reduced significantly during the perturbation due to the correspondingly narrowed soliton bandwidth, as illustrated in Fig. S3(b). For the case with a vanishing soliton, the damping strength could become negligible, and the soliton would keep oscillating even when it has almost disappeared in the noise background of the laser field. Nevertheless, comprehensive understanding of the damping mechanism would require further investigation, while the dissipative model provided in this work can serve as a simple guide to understand the dominant factors in the retiming dynamics of the solitons in the acoustic lattice.